\documentclass[prb,twocolumn,showpacs,preprintnumbers,superscriptaddress,groupedaddress]{revtex4-1}

\usepackage{amsmath}
\usepackage{amssymb}
\usepackage{graphicx}

\begin{document}

\title{Interplay between periodicity and nonlinearity of indirect excitons in coupled
quantum wells}

\author{T. F. Xu$^{1}$, X. L. Jing$^{1}$, H. G. Luo$^{2,3}$, W. C. Wu$^{4}$, and C. S. Liu$^{1,4}$}

\affiliation{$^1$Department of Physics, Yanshan University, Qinhuangdao 066004, China\\
$^2$Center for Interdisciplinary Studies and Key Laboratory for Magnetism and
Magnetic Materials of the Ministry of Education, Lanzhou University, Lanzhou 730000, China\\
$^3$Beijing Computational Science Research Center, Beijing 100084, China\\
$^4$Department of Physics, National Taiwan Normal University, Taipei 11677, Taiwan}

\date{\today}

\begin{abstract}
Inspired by a recent experiment of localization-delocalization transition (LDT)
of indirect excitons in lateral electrostatic lattices [M. Remeika \textit{et al.},
Phys. Rev. Lett. \textbf{102}, 186803 (2009)], we investigate the interplay
between periodic potential and nonlinear interactions of indirect excitons
in coupled quantum wells. It is shown that the model involving both attractive
two-body and repulsive three-body interactions can lead to a natural account for the LDT
of excitons across the lattice when reducing lattice amplitude or increasing
particle density. In addition, the observations that the smooth component of
the photoluminescent energy increases with increasing exciton density and
exciton interaction energy is close to the lattice amplitude at the transition
are also qualitatively explained. Our model provides
an alternative way for understanding the underlying physics of the exciton
dynamics in lattice potential wells.
\end{abstract}

\maketitle

\section{Introduction}

Indirect excitons (spatially separated electron-hole pairs) in coupled quantum
wells (CQWs) can have a long lifetime and a high cooling rate. With these two merits,
Butov \textit{et al.} have successfully cooled the trapped excitons to the order of 1K
and have observed surprisedly that excitons form two, inner and external, rings to which
periodic bright spots appear in the external ring. \cite{Butov2002a,Butov2002b}
Although there is no clear evidence that these excitons are condensed into the BEC state,
it is fascinating enough to see some puzzling particle number
distributions in various confined potential wells. \cite{LaiCW2004, hammack:227402}
More recently, periodic potential (lattices) due to gate voltages
was created for indirect excitons of built-in electric dipole moment.
It enables observation of localization-delocalization transition (LDT)
for transport across the lattice with reducing lattice amplitude or increasing
particle density.\cite{PhysRevLett.102.186803} This gives an opportunity to
understand further the complex dynamical behaviors of indirect excitons.

In the literature, a charge separated transportation mechanism was proposed,
\cite{butov:117404} which gave a satisfactory explanation to the formation of
exciton rings and the dark region between the inner and the external rings.
However the origin of the periodic bright spots in the external ring can not
be fully understood within the above framework. Alternatively a self-trapped
interaction model involving attractive two-body and repulsive three-body
interactions in the system was proposed. \cite{Liu2006, PhysRevB.80.125317}
It was shown that the interplay between the two-body attraction and the three-body
repulsion can give a good account to the periodic bright spots in the external ring.
In addition, the self-trapped interaction model also explained well the abnormal
exciton distribution in an impurity potential, where the photoluminescence (PL)
pattern becomes much more compact than a Gaussian with a central intensity dip,
exhibiting an annular shape with a darker central region. \cite{LaiCW2004}
Moreover, the model also captured some other experimental details, for instances,
the dip can turn into a tip at the center of the annular cloud when the sample
is excited by higher power lasers.

The success of the self-trapped interaction model in understanding
various phenomena has motivated us to investigate the LDT phenomenon reported in
Ref.~[\onlinecite{PhysRevLett.102.186803}] using the same model. It will be
shown that the complex LDT as well as other dynamic behaviors can also
be qualitatively explained by the self-trapped interaction model.

This paper is organized as follows. In Sec.~\ref{Model}, we introduce the
self-trapped interaction model described by a phenomenological nonlinear
Schr\"{o}dinger equation. An important factor regarding
how the exciton distribution depends on temperatures and energies is detailed.
In Sec.~\ref{Results}, numerical calculations and detailed discussions are
given for studying the LDT of transports across the lattice.
It will be shown that the model introduced in Sec.~\ref{Model} can lead
to a reasonably good explanation for LDT.
Sec.~\ref{Summary} is a brief summary.

\section{Model} \label{Model}

The idea of the self-trapping model came from the density dependence of various observed
PL spectroscopy. \cite{Butov2002b, LaiCW2004, hammack:227402, PhysRevLett.102.186803}
As a matter of fact, intensity of PL spectroscopy is proportional to the exciton
number density and therefore complex experimental PL data will reveal complex
interaction between excitons. It is clear that the interaction between excitons is
neither purely attractive, nor purely repulsive. If it is purely repulsive, it could
drive the system towards homogeneous distribution. On the contrary, if it's purely
attractive, the system will become unstable and eventually collapse
when the exciton density is larger than some critical value.
Thus it is reasonable to speculate that at low densities the interaction
is dominated by an attraction, while at high densities a repulsion
must exist to prevent the system against collapses. \cite{Liu2006}
The observation that exciton cloud first contracts then expands by increasing the
excitation power gives a strong support to the above scenario. \cite{LaiCW2004}

Excitons behave as electric dipoles, so strong Coulomb repulsion will govern the
indirect excitons in CQWs at high densities to which dipoles are aligned
in parallel. \cite{Butov1990}
However at low densities such that two dipoles change from aligning in parallel
to inclined, the attraction between the electron of one exciton and the
hole of the other exciton will dominate instead.\cite{sugakov:115303}
In addition to Coulomb interactions, exchange effect may be another
important factor for interactions. When two excitons approach to
each other, exchange interaction between two electrons (or two holes)
will become more important. This may be another source of the attraction
between excitons.

In the dilute limit, it is assumed that indirect excitons in CQWs are governed
by two-body attractions and three-body repulsions. Equivalently one can consider
the following scaled \emph{nonlinear} Schr\"{o}dinger equation
\cite{Liu2006, PhysRevB.80.125317}
\begin{equation}
-\frac{1}{2}\nabla^{2}\psi_{j}({\bf r})+[V_{\rm ex}({\bf r})-g_{1}n+g_{2}n^{2}]\psi
_{j}({\bf r})=E_{j}\psi_{j}({\bf r}),
\label{the schrodinger equation}
\end{equation}
where $\psi_{j}(\mathbf{r})$ and $E_{j}$ are the \textit{j}-th energy eigenfunction
and eigenvalue and $V_{\rm ex}$ denotes the external potential.
The eigenfunction is normalized under $\int_\Omega |\psi_{j}(\mathbf{r} )|^{2}d\Omega=1$.
In Eq.~(\ref{the schrodinger equation}), all lengths and energies are scaled in
units of $\sigma_{\rm PL}$ and $\hbar^2/m^*\sigma^2_{\rm PL}$ respectively,
where $m^*=m_e/2$ is effective mass of the exciton and $\sigma_{\rm PL}$ is the
root-mean-square radius of the exciton cloud observed via PL. \cite{PhysRevB.80.125317}
Considering the special property of the system, local probability density of excitons
[$n=n(\mathbf{r})$] will be taken to be
\begin{equation}
n(\mathbf{r}) = \sum_{j=1}^{\mathcal{N}} \eta(E_{j})
|\psi_{j}(\mathbf{r} )|^{2},\label{eq:density}
\end{equation}
where $\mathcal{N}$ denotes the total number of bound
states involved and $\eta(E_{j})$ is the probability (or distribution) function
for energy-level $E_{j}$ that satisfies the normalization condition
$\sum_{j=1}^{\mathcal{N}}\eta(E_{j})=1$. At the mean-field level,
two- and three-body coupling constants, $g_{1}$ and $g_{2}$, are defined positively,
which will be proportional to $N$ and $N^2$ with $N$ being the number of excitons.
Note that exciton number density can be controlled by and proportional to the
excitation laser power $P$, thus $g_1$ is proportional to $P$ in the current model.

Determination of the distribution function $\eta(E_{j})$ in (\ref{eq:density})
requires considering both complex {\em energy relaxation} and {\em recombination}
processes. At low exciton densities ($n\ll 1/a_{B}^{2}$, $a_{B}$ being Bohr radius),
relaxation due to exciton-exciton and exciton-carrier scattering can
be neglected \cite{Piermarocchi1996} and consequently
relaxation time is dominated by the scattering of excitons with acoustic phonons.
At low bath temperatures, $(T_{b}<1$ $\mathrm{K})$, this kind of relaxation rate
decreases dramatically due to the low phonon density. \cite{Benisty1991}
For the recombination process, on the other hand, because excitons in
the lowest self-trapped level are quantum degenerate, they are
dominated by the stimulated scattering when the occupation number is
more than a critical value. Strong enhancement of the exciton
scattering rate has been observed in the resonantly excited
time-resolved PL experiment. \cite{butov:5608} Therefore, even
though the phonon scattering rate is larger than the radiative
recombination rate, it is in reality that the system does not
reach thermal equilibrium. Following Ref.~[\onlinecite{PhysRevB.80.125317}],
in accord with the discrete PL spectroscopy reported in Ref.~[\onlinecite{high-2008}],
$\eta(E_{j})$ will have the following form
\begin{eqnarray}
\eta(E_j) = C\exp[\alpha(E_j-\mu)],
\label{eq:eta}
\end{eqnarray}
where $C$ is the normalization factor, $\mu$ is the chemical potential, and
$\alpha$ can be viewed as an ``effective temperature".

\begin{figure}[tbp]
\begin{center}
\includegraphics[width=7.5cm]{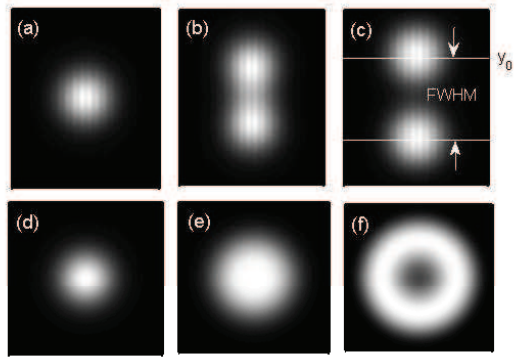}
\end{center}\vspace{-0.8cm}
\caption{Calculated spatial probability density distribution of highly degenerate
indirect excitons, $n(x,y)$, for low densities ($g_1=10$) [frame (a) \& (d)],
intermediate densities ($g_1=20$) [frame (b) \& (e)], and high densities
($g_1=30$) [frame (c) \& (f)]. Frame (a)-(c) are with a lattice potential
$V_{\rm ex}=v_0\cos(2\pi x/\lambda)$ ($v_0=50$ and $\lambda=0.05$),
while (d)-(f) are without a lattice potential ($v_0=0$). In all frames,
$g_{2}\equiv 0.001g_{1}^{2}$, $\alpha$ is fixed at $0.1$, and $x$ and $y$ axes are
in units of $\sigma_{\rm PL}$ (see text).
The results are intended to be compared to those in Fig.~1
of Ref.~\onlinecite{PhysRevLett.102.186803}. In frame (c), FWHM (calculated
in Fig.~3) and $y_0$ (used in Fig.~2) are denoted.}
\label{fig1}
\end{figure}

\section{Results and Discussions} \label{Results}

\subsection{Spatial exciton distributions with or without lattices}

Key features of indirect excitons with or without lattices, as reported in
Ref.~[\onlinecite{PhysRevLett.102.186803}], are the following.
LDT for transport across the lattice was observed with reducing lattice amplitude
or increasing exciton density.
The inner ring, which was observed in the PL patterns of indirect excitons
without lattices, persists in the lattice case. For relatively low excitation
power ($P = 3.7$ $\mu$W), outer ring effect does not emerge because of
low particle number density and hence low drift and diffusion of
electrons and holes. However, for relatively higher excitation power ($P = 12$ $\mu$W),
in addition to the inner ring, outer ring seems to exist in the case without the lattices
(see Fig.~1.(i) of Ref.~[\onlinecite{PhysRevLett.102.186803}]).
With the lattices (see Fig.~1.(f) of Ref.~[\onlinecite{PhysRevLett.102.186803}]),
outer ring transforms to spatially separated clouds, while inner ring persists.


Based on the above observations, in the current localized-delocalized system it seems
that there are two kinds of excitons, whose spatial distributions could be governed
by different mechanisms. The ones associated with the inner ring effect are ascribed
to the {\em hot} excitons which have been studied in great details in
Refs.~\cite{Butov2002b, butov:117404, rapaport:117405,PhysRevLett.102.186803}.
The other ones are ascribed to the {\em cool} excitons whose formation and
pattern are considered to be due to quantum degeneracy.\cite{Liu2006, PhysRevB.80.125317}.

When excitation power $P$ is low, densities of both hot and cool excitons are low.
They are all localized within the range of the laser spot (i.e., localization).
When $P$ is increased, densities of both kinds of excitons will increase
for which inner ring forms due to transportation and cooling of the hot excitons.
While cool excitons will form patterns due to quantum degeneracy arising from
complex interactions. Delocalization effect increases with increasing the laser
power $P$, to which radius of the inner ring as well as the distance between the
two separated degenerate exciton clouds increase.
As a matter of fact, full width at half maximum (FWHM) of the exciton cloud will
change with the particle number density or the lattice potential amplitude, as
shown in Fig.~3 of Ref.~\onlinecite{PhysRevLett.102.186803}.

Here we shall focus on the LDT associated with highly degenerate low-energy (cool)
excitons only. Spatial distributions of low-energy exciton density are calculated
self-consistently based on Eqs.~(\ref{the schrodinger equation})--(\ref{eq:eta}).
The results are shown in Figs.~\ref{fig1}(a)--\ref{fig1}(c) for the case of a periodic lattice
potential, $V_{\rm ex}=v_0\cos(2\pi x/\lambda)$ with $v_0=120$ and $\lambda=0.05$,
and in Figs.~\ref{fig1}(d)--\ref{fig1}(f) for the case without a lattice potential ($v_0=0$).
In the case with lattice potential, a weak rectangular potential well is applied
to ensure the strip pattern along the lattices ({\em i.e.}, with the Dirichlet
boundary condition). In our model, highly degenerate excitons are self-trapped due to
competition between complex interactions and the nonequilibrium energy distribution.
Apart from the periodic lattice potential,
any additional weak external potential will only have minor effect on the results.
In our calculations, the wave function is chosen to be a Gaussian initially.
All self-trapped energy eigenstates $\psi_{j}$ with $\varepsilon_{j}<0$ are
solved and will contribute to the spatial distribution.

As mentioned before, at low excitation powers $P$ low-energy exciton density profile
essentially coincides with the excitation laser spot. In our model, this arises
due to the dominant attraction between excitons in the dilute limit and
at the same time only ground state contributes to the self-trapping. The ground-state
wave function is $s$-wave with a maximum at the center [see Figs.~\ref{fig1}(a) and
\ref{fig1}(d)]. Excitons are localized and do not travel beyond the excitation spot.

When exciton number is increased with higher excitation power $P$, excitons can
delocalize and spread beyond the excitation spot [see Figs.~\ref{fig1}(b), \ref{fig1}(c),
\ref{fig1}(e), and (\ref{fig1}f)]. In these cases, effect of repulsion becomes more
important which results more energy eigenstates involved in the self-trapping.
The first excited states are nearly two-fold degenerate $p$-wave with a node at the center.
Superposition of the ground-state $s$-wave and the two first-excited-state $p$-wave
wavefunctions leads to an annular distribution with a hole at the center
in the case without lattices [see Fig.~\ref{fig1}(f)]. With the lattices,
two-fold degeneracy is lifted and consequently ring-shaped delocalized excitons
shift to two separate clouds at high excitation powers [see Fig.~\ref{fig1}(c)].

\begin{figure}[tbp]
\begin{center}
\includegraphics[width=8.5cm]{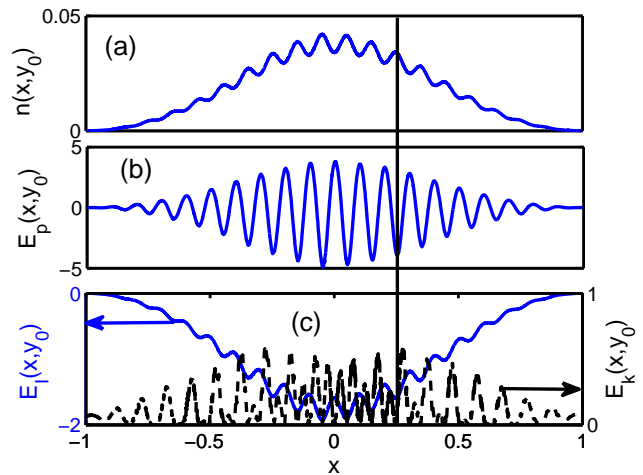}
\end{center}\vspace{-0.6cm}
\caption{(Color online) (a) Exciton probability density distribution $n(x,y_0)$
with $y_0$ denoted in Fig.~\ref{fig1}(c). (b) Lattice potential energy density
$E_p(x,y_0)\equiv V_{\rm ex}(x)n(x,y_0)$ plotted vs. $x$. (c) Spatially dependent
interaction energy density $E_{I}(x,y_0)$ and kinetic energy density $E_{k}(x,y_0)$
plotted vs. $x$. Parameters used are the same as those in Fig.~\ref{fig1}(c),
except a larger $g_1=50$ and $v_0=100$ to account for the experiment. Vertical line shows
that maxima in density $n(x,y_0)$ correspond to minima in periodic
potential energy density $E_p(x,y_0)$. In all frames,
$x$ axis is in units of $\sigma_{\rm PL}$. The results are intended to be
compared to those in Fig.~2 of Ref.~\onlinecite{PhysRevLett.102.186803}.}
\label{fig2}
\end{figure}

\subsection{Spatially dependent PL energy}

A PL image of indirect excitons in the delocalized regime, plotted as energy vs.
space, is presented in Fig.~2(a) of Ref.~\onlinecite{PhysRevLett.102.186803}.
As a matter of fact, both integrated PL intensity $I(x)$ and the average PL energy
$\hbar\omega(x)$ show a small modulation at the lattice period superimposed on
a smoothly varying profile. In particular, the average PL energy shows a dome carve shape.

In fact in CQWs, spatially dependent energy of indirect excitons, which is directly
related to the PL energy $\hbar\omega$, involves four major contributions.
The first part is the intrinsic energy that includes contributions due to
the band gap, the Coulomb interaction between electrons and holes,
and the electric potential used to form indirect excitons. Intrinsic energy
of exciton mainly contributes to the smooth part of the PL energy $\hbar\omega$
since it does not change with positions when experimental condition remains unchanged.
The second part is the kinetic energy $E_k$. Near the laser spot, excitons are
hot with a large kinetic energy. It is believed that this is the main cause
why a high energy distribution is located in the center regime. The third part
is the interaction energy $E_I$. It also contributes to the smooth part of the
PL energy $\hbar\omega$. Within our model, when exciton interaction is in
the repulsion-dominant regime, $E_I$ increases with the exciton
density $n$ and thus smooth part of $\hbar\omega$ will increase with increasing
the excitation power $P$. Moreover, reducing the cloud size corresponds to
increasing the exciton density when excitation power $P$ remains unchanged.
Therefore smooth part of $\hbar\omega$ will increase with reducing the cloud size.
The last part is the lattice potential energy $E_p$.
Within the Thomas-Fermi approximation, it is easy to verify that $E_p$
dominates over $E_I$ and $E_k$ in such a dilute system (see later).
This is why maxima in exciton number density (and equivalently the PL intensity)
correspond to minima in the periodic potential energy density $E_p$ (see Fig.~2).

Within the same model, average energy density
and number distribution of highly degenerate excitons are calculated and
shown in Fig.~\ref{fig2}. These are intended to be compared to those
reported in Fig.~2 of Ref.~\onlinecite{PhysRevLett.102.186803}.
Apart from the smooth part mainly due to the intrinsic energy,
spatially dependent kinetic energy density of degenerate excitons
can be calculated via\cite{PhysRevB.80.125317}
\begin{equation}
E_{k}({\bf r})=\sum\limits_{j=1}^{\mathcal{N}}\left[ \eta
(E_{j})\left( \left\vert \frac{\partial \psi _{j}({\bf r})}{\partial
x}\right\vert ^{2}+\left\vert \frac{\partial \psi
_{j}({\bf r})}{\partial y}\right\vert ^{2}\right) \right],
\label{Ek}
\end{equation}
while the interaction (mean-field) energy density is given by
$E_{I}({\bf r})=-g_1 n({\bf r})^2+g_2 n({\bf r})^3$ and the lattice potential
energy density is given by $E_{p}({\bf r})=V_{\rm ex}({\bf r})n({\bf r})$,
which are obtained by multiplying $\eta(E_j)\psi _{j}^*({\bf r})$ to
Eq.~(\ref{the schrodinger equation}) and summing over $j$.

Fig.~\ref{fig2}(a) shows the calculated spatial probability density distribution
of degenerate excitons, $n(x,y_0)$. Here $y_0$ corresponds to center of the upper
cloud in $y$ direction [see Fig.~\ref{fig1}(c)]. The spatial lattice potential
energy density $E_p(x,y_0)$, kinetic energy density $E_{k}(x,y_0)$, and interaction
energy density $E_{I}(x,y_0)$ are presented in Figs.~\ref{fig2}(b) and \ref{fig2}(c)
respectively. In view of Fig.~\ref{fig2}, $E_{k}$ and $E_I$ are about one order of
magnitude smaller than $E_p$. This means that Thomas-Fermi approximation is valid
in such a dilute exciton system. Consequently exciton number distribution is mainly
determined by the periodic potential energy $E_p$. The higher the $E_p$ is,
the larger the local PL energy is. On the contrary, the lower the $E_p$ is,
the higher the local exciton number density is. As PL intensity is proportional
to the local exciton number density, it results that minima in energy correspond
to the maxima in intensity. Equivalently the largest exciton number density (and
hence the PL intensity) are located in the bottom of the periodic potential energy.

\begin{figure}[tbp]
\begin{center}
\includegraphics[width=8.5cm]{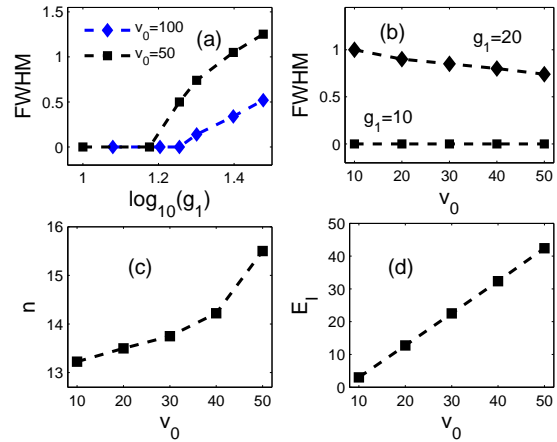}
\end{center}
\vspace{-0.6cm}
\caption{(Color online) Calculated FWHM of delocalized exciton clouds vs. (a)
$\log_{10}(g_1)$ ($g_1\propto$ exciton number density $n$) for lattice amplitude
$v_0=50$ and $100$ and (b) $v_0$ for different $n$ with $g_1=10$ and $20$.
(c) Calculated number density $n$ at the LDT point as a function of lattice
amplitude $v_0$. (d) Calculated interaction energy density $E_I(0,y_0)$ at the
LDT point as a function of lattice amplitude $v_0$. $(0,y_0)$ corresponds to
the outer cloud center. The results are intended to be compared to Fig.~3
in Ref.~\onlinecite{PhysRevLett.102.186803}.}
\label{fig3}
\end{figure}

\subsection{Interplay between periodic potential and nonlinear interaction}

In view of the results shown in Figs.~3(a) and 3(b) of
Ref.~[\onlinecite{PhysRevLett.102.186803}], it is surprising to see that
FWHM of the exciton cloud (across the lattice) decreases with increasing the
lattice amplitude. On the contrary, the excitation power needed for the transition
from localized to delocalized regimes increases with the increase of the lattice
amplitudes [see Figs.~3(c) of Ref.~\onlinecite{PhysRevLett.102.186803}].
This is likely due to the reduction of the particle tunneling rate across
the lattice. Thus a large laser power is needed to obtain exciton delocalized
distribution when increasing the lattice amplitude.

Fig.~\ref{fig3} studies the effects of number density and lattice amplitude
on LDT of degenerate excitons. Here, as the dynamics of outer cloud is focused on,
we define FWHM as the distance (the extension) between centers of the two clouds along the
$y$-direction [see Fig.~\ref{fig1}(c)]. In Figs.~\ref{fig3}(a) and \ref{fig3}(b),
we show FWHM as a function of the particle number density (via
changing $g_1$) and the lattice potential amplitude respectively.
For weak excitation power $P$ and low exciton number correspondingly,
within our model both the attractive and repulsive interactions
between excitons are weak. All particles are sitting in the
ground state and self-trapped in the shallow mean-field potential well.
In this case, FWHM of exciton cloud is equal to zero and
remains unchanged in the lower excitation power regime
[see Fig.~\ref{fig3}(a) and Fig.~\ref{fig3}(b) with $g_1$=10].
When the excitation power $P$ or the particle density $n$ is increased to above some
critical point, LDT occurs. Within our model, repulsive interaction
dominates and consequently excitons will be sitting in both ground state and two-fold
degenerate first-excited states. Due to the one-dimensional lattice potential,
two-fold degeneracy of excited states is lifted and consequently two separated
delocalized exciton clouds form.

In Fig.~\ref{fig3}(c), we show the critical number density $n$ for LDT
as a function of lattice amplitude $v_0$. When $v_0$ is increased, before
the LDT transition excitons are strongly localized in a single well center,
while particle tunneling rate across the well is largely decreased.
As a consequence, critical number density for LDT increases.
Within our model, before the LDT transition a larger lattice amplitude $v_0$
corresponds effectively (or relatively) to a smaller exciton density.
This also means that the effect of the repulsion is decreased while
the effect of the attraction is increased. This explains why a
blue-shift of the LDT is observed in the experiment.
It gives another evidence to support our model.

In the delocalized regime, when excitation power $P$ is low, only intrinsic
energy and potential energy contribute to the exciton PL energy because
interaction energy $E_I$ is immaterial. When $P$ is increases, however, $E_I$ will
become more important and contribute much more to the PL energy.
Therefore, $\hbar\Delta\omega$, defined as the difference between $\hbar\omega$
and that associated with the lowest $P$ for LDT, is dominated by $E_I$.
Within our model, in the delocalized repulsion-dominant regime, interaction energy
density $E_I$ will increase monotonically with increasing the lattice amplitude $v_0$.
Fig.~\ref{fig3}(d) shows the calculated interaction energy density $E_I(0,y_0)$ at the
LDT point as a function of lattice amplitude $v_0$. Here
$(x,y)=(0,y_0)$ corresponds to the center of the outer cloud.
In agreement with the experiment, it is seen that $E_I$ at the LDT point
is close to the value of the lattice amplitude $v_0$.
This agreement gives a strong support that the scaled parameters used in the
current context are reasonable.

\section{Summary}\label{Summary}

In summary, we have investigated the interplay between periodic potential
and nonlinear interaction of indirect excitons in coupled quantum wells, with
or without a lattice potential. The model which takes into account the
competition between a two-body attraction and a three-body repulsion along
with a reasonable nonequilibrium energy distribution gives an alternative
qualitatively good account on the localization-delocalization transitions (LDT)
of excitons across the lattice when increasing the particle density or
reducing the lattice amplitude.

\begin{acknowledgments}
This work was supported by National Science Council of Taiwan
(Grant No.99-2112-M-003-006), Hebei Provincial Natural Science Foundation of China (Grant
No.A2010001116 and D2010001150), and National Natural Science
Foundation of China (Grant No.10974169 and 10874059). We also acknowledge the
support from NCTS, Taiwan.
\end{acknowledgments}

%

\end{document}